\documentclass{aastex631}

\def\apjl{ApJL}
\def\apjs{ApJS}
\def\aap{A\&A}

\def\be{\begin{equation}} 
\def\ee{\end{equation}} 
\def\ba{\begin{eqnarray}} 
\def\ea{\end{eqnarray}}

\def\msun{{\Msun}}

\def\HH{${\rm {H_2}}$}

  \def\cm{${\rm cm}$}

\def\gsim{\lower.5ex\hbox{\gtsima}} 
\def\lsim{\lower.5ex\hbox{\ltsima}} \def\gtsima{$\; \buildrel > \over 
\sim \;$} \def\ltsima{$\; \buildrel < \over \sim \;$} \def\prosima{$\; 
\buildrel \propto \over \sim \;$} \def\gsim{\lower.5ex\hbox{\gtsima}} 
\def\lsim{\lower.5ex\hbox{\ltsima}} 
\def\simgt{\lower.5ex\hbox{\gtsima}} 
\def\simlt{\lower.5ex\hbox{\ltsima}} 
\def\simpr{\lower.5ex\hbox{\prosima}}   
  
 \def\gtsima{$\; \buildrel > \over \sim \;$} 
\def\ltsima{$\; \buildrel < \over \sim \;$} 
\def\gsim{\lower.5ex\hbox{\gtsima}} 
\def\lsim{\lower.5ex\hbox{\ltsima}} 
\def\simgt{\lower.5ex\hbox{\gtsima}} 
\def\simlt{\lower.5ex\hbox{\ltsima}} 
\def\simpr{\lower.5ex\hbox{\prosima}}

\def\msun{\,{\rm \Msun}}

\def\E3{{\cal E}_{\rm g}^{III}}

\def\r12{r_{1/2}} 
\def\x12{x_{1/2}} 
\def\v12{v_{1/2}}

\newcommand\code[1]{\textsc{\MakeLowercase{#1}}}

\def\nh2{n_{\rm H2}}
\def\fh2{f_{\rm H2}}

\def\arcsec{^{\prime\prime}}
\def\angstrom{\textrm{A\kern -1.3ex\raisebox{0.6ex}{$^\circ$}}}
\def\myr{\rm Myr}

\def\msun{{\rm M}_{\odot}}
\def\zsun{{\rm Z}_{\odot}}

\def\msunyr{\msun\,{\rm yr}^{-1}}

\def\kpc{{\rm kpc}}

\def\althaea{Alth{\ae}a}
\def\highz{$\mbox{high-}z$~}

\makeatletter
\def\@hex@@Hex#1%
 {\if a#1A\else \if b#1B\else \if c#1C\else \if d#1D\else
  \if e#1E\else \if f#1F\else #1\fi\fi\fi\fi\fi\fi \@hex@Hex}
\makeatother

\definecolor{apcolor}{HTML}{b3003b}
\definecolor{cbcolor}{HTML}{ff0f00}
\definecolor{afcolor}{HTML}{b3443c}
\definecolor{ddcolor}{HTML}{077a2f}
\definecolor{vgcolor}{HTML}{8F00FF}

\usepackage[english]{babel}
\usepackage{amsmath}
\usepackage{amssymb}
\usepackage{graphicx}
\usepackage{subfigure}
\usepackage[normalem]{ulem}

\usepackage{verbatim}

\usepackage{color}
\usepackage{multirow}
\usepackage{mathtools}
\usepackage{epstopdf}

\usepackage{url}
\usepackage{xcolor}

\shorttitle{LBG dwarf satellites with JWST}
\shortauthors{Gelli et al.}

\begin{document}

\title{Dwarf satellites of high-$z$ Lyman Break Galaxies: a free lunch for JWST}

\correspondingauthor{Viola Gelli}
\email{viola.gelli@unifi.it}

\author[0000-0001-5487-0392]{Viola Gelli}
\author[0000-0001-7298-2478]{Stefania Salvadori}
\affiliation{Dipartimento di Fisica e Astronomia, Universit\'{a} degli Studi di Firenze, via G. Sansone 1, 50019, Sesto Fiorentino, Italy}
\affiliation{INAF/Osservatorio Astrofisico di Arcetri, Largo E. Fermi 5, I-50125, Firenze, Italy}
\author[0000-0002-9400-7312]{Andrea Ferrara}
\author[0000-0002-7129-5761]{Andrea Pallottini}
\author[0000-0002-6719-380X]{Stefano Carniani}
\affiliation{Scuola Normale Superiore, Piazza dei Cavalieri 7, I-56126 Pisa, Italy}

\begin{abstract}
We show that the \textit{James Webb Space Telescope} will be able to detect dwarf satellites of \highz Lyman Break Galaxies (LBGs). To this aim, we use cosmological simulations following the evolution of a typical  $M_\star\simeq10^{10}\rm \msun$ LBG up to $z\simeq6$, and analyse the observational properties of its five satellite dwarf galaxies ($10^7{\rm \msun}<M_\star<10^9{\rm \msun}$). 
Modelling their stellar emission and dust attenuation, we reconstruct their rest-frame UV-optical spectra for $6<z<6.5$. JWST/NIRCam synthetic images show that the satellites can be spatially resolved from their host, and their emission is detectable by planned deep surveys. Moreover, we build synthetic spectral energy distributions and colour-magnitude diagrams for the satellites. We conclude that the color F200W-F356W is a powerful diagnostic tool for understanding their physical properties once they have been identified. For example, F200W-F356W$\lesssim-0.25$ can be used to identify star-bursting (${\rm {\rm SFR}}\sim5~\msunyr$), low-mass ($M_\star\lesssim5\times 10^8\rm \msun$) systems, with $\sim80\%$ of their stars being young and metal-poor [$\log(Z_\star/Z_\odot) < -0.5$]. 
\end{abstract}

\keywords{High-redshift galaxies --- Dwarf galaxies --- Cosmology}

\section{Introduction}%%%%%%%%%%%%%%%%%%%%%%%%%%%%%%%%%%%%%%

The James Webb Space Telescope \citep[JWST,][]{Gardner06} will provide unprecedented observations of the most distant galaxies, revolutionising our knowledge of the high redshift Universe. It will be able to detect remarkably faint objects during the Epoch of Reionization (EoR, $z\gtrsim6$) reaching UV absolute magnitudes of $\rm M_{UV}\simeq-15$ at redshift $z\simeq6$ \citep[e.g.][]{Finkelstein16}. The wavelength coverage of instruments like NIRCam \citep[$0.6-5\, \micron$,][]{HornerRieke04} will allow us to study their rest-frame optical and UV properties for the first time.

Among the faint objects expected to be present in large numbers at \highz are dwarf galaxies ($M_\star\lesssim10^9\msun$). Indeed, according to the hierarchical $\rm \Lambda CDM$ model, dwarf galaxies are the most abundant systems at all cosmic times, being the first galaxies to form and the basic building blocks of the massive galaxies we see today.

The physical and observational properties of \textit{isolated} dwarf galaxies located in the field have been widely studied through cosmological simulations \citep{Wise14,Zackrisson17, JeonBromm19}. 
However, we also expect the presence of an additional category of dwarf galaxies: the \textit{satellites} orbiting around more massive galaxies in the densest regions of the cosmic web \citep[e.g][]{Conroy06}.
In particular, \citet{Gelli20} show that a typical massive $z\simeq 6$ Lyman Break Galaxy (LBG) is surrounded by dwarf satellites, living inside its virial halo. These satellites represent an interesting complementary population with respect to isolated field dwarf galaxies, since their properties and evolution are affected by the peculiar environment in which they dwell.

JWST is expected to detect many LBGs: the emission of dozens of them will most likely appear already within the deep fields of high priority surveys like JADES \citep{Rieke19}, that will reach exposure times of $\sim20$~hrs for imaging with NIRCam.
This provides the remarkable opportunity to exploit planned JWST surveys to detect {\it for the first time} and {\it for free} their dwarf satellite population.

Will JWST be able to effectively detect dwarf satellites of massive LBGs through imaging and photometry? Will such observations shed light on the star formation rates, stellar metallicity and presence of young/metal-poor stars in these systems?
We address these issues by using high-resolution, zoom-in cosmological simulations \citep{Pallottini17} reproducing the evolution of a typical $z\sim6$ LBG. In \citet{Gelli20} we studied the stellar populations of the dwarf satellites. Here, we make a further step and model their emission, focusing on the stellar continuum. 

\section{Method}%%%%%%%%%%%%%%%%%%%%%%%%%%%%%%%%%%%%%%
\subsection{Simulating high-$z$ LBGs and their dwarf satellites}
To study the emission of LBGs dwarf satellites at high-$z$, we use a high resolution ($\lesssim 30$~pc) zoom-in cosmological simulation, fully described in \citet{Pallottini17}.
It adopts a customised version of the adaptive mesh refinement code \code{RAMSES} \citep{Teyssier02} to follow the evolution of a proto-typical LBG called \althaea.
Stellar particles in the simulation form according to an \HH~dependent Schmidt-Kennicutt relation \citep{Kennicutt98}, where the abundance of molecular hydrogen is computed by a non-equilibrium chemical network via the \code{KROME} code \citep{krome}.
Stellar evolution is modelled accounting for energy inputs and yields by using \code{STARBURST99} \citep{starburst99}, assuming \code{padova} \citep{Bertelli94} stellar tracks and a Kroupa initial mass function \citep{Kroupa01}.
Since the simulation cannot follow the formation of the first stars in pristine mini-halos, a metallicity floor of $Z_{floor}=10^{-3}\zsun$ has been imposed to mimic the consequences of the first stellar generations on metal enrichment \citep{Wise12,Pallottini14}.

At $z\sim6$ the simulated LBG \althaea~has a stellar mass of $M_\star \sim 10^{10} \rm  M_\odot$ and a star formation rate $\rm SFR\sim50~\msunyr$.

The target halo hosts five dwarf satellites and one proto-globular cluster orbiting around the central massive LBG ($r<10~\kpc$); their stellar properties are thoroughly analysed in \citet{Gelli20}.
The satellites can be divided into two main groups based on their mass, differing in terms of both evolutionary and chemical properties.
Low-mass systems ($M_\star \lesssim 5 \times 10^8\msun$) form their stellar populations in short bursts ($\sim 50$~Myr) after which star formation is completely suppressed by supernovae (SN) feedback. More massive satellites are characterised instead by long and complex star formation histories \citep[SFH, see Fig.~7 and 8 of][]{Gelli20}, resulting from the balance between feedback and merger events bringing new gas to fuel star formation.

Perhaps surprisingly, at $z\sim6.5$ the interstellar medium of all the satellites is already enriched in heavy elements ($Z_{\rm ISM} \geq 10^{-2} \zsun$), and most likely in dust. 
These high metallicities are due to: (i) pre-enrichment of the environment by the central LBG and its progenitors, for the case of low-mass satellites; (ii) in-situ star formation for massive satellites, which form their first stellar generations at $z>9$ in a nearly pristine environment \citep{Gelli20}.
In spite of their low initial metallicities, high-mass satellites contain a smaller fraction of metal-poor (defined as [$\log(Z_\star/Z_\odot) < -0.5$]) stars ($\lesssim20\%$) with respect to smaller systems ($\gtrsim 50\%$) because of their longer SFHs.

\subsection{SED building and synthetic images}

To derive the intrinsic spectral energy distributions (SED) of the dwarf satellites based on their simulated stellar populations, we use \code{STARBURST99} \citep{starburst99}, also consistent with the simulation prescriptions.
We proceed as follows: considering the total SFH of the selected galaxy (up to $z\sim6$, final snapshot of the simulation), we sample it in timesteps of ${\rm \Delta} t = 5~\myr$, which assure a good convergence for the spectra. 
At each timestep we consider a single burst of star formation: the stellar mass, metallicity and formation time are given as an input to \code{STARBURST99}. The final output consists in the rest-frame intrinsic spectrum of the galaxy at all times from its formation up to $z\sim6$.

\begin{figure*}
\centering
\includegraphics[width=\textwidth]{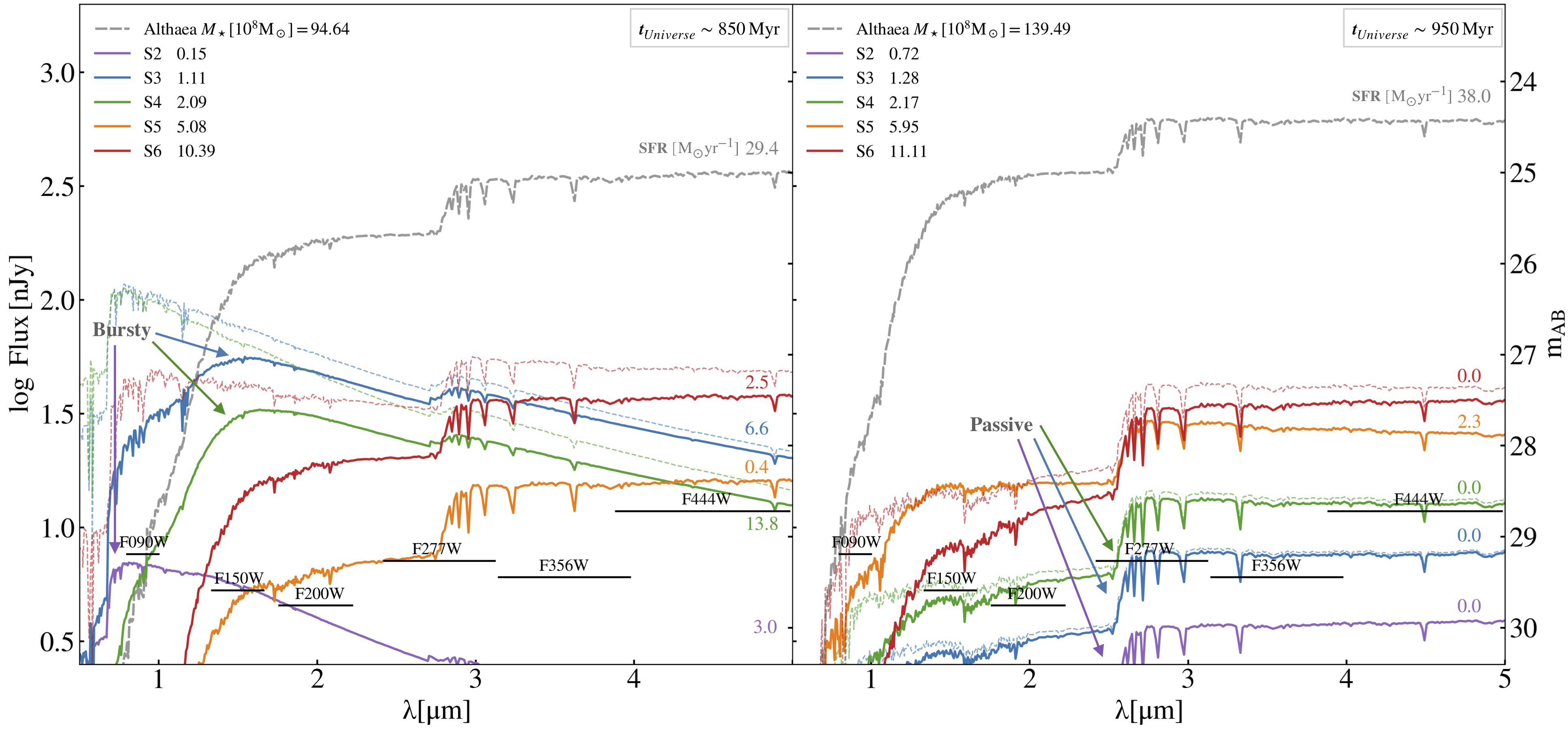}
\caption{Spectral energy distributions for the five \highz satellites (coloured continuous curves) and the central LBG Althaea (grey dashed) shown at two different times. \textit{Left:} \textit{Bursty phase} ($z\simeq6.5$), in which the three smallest satellites are experiencing a star formation burst. \textit{Right:} \textit{Passive phase} about 100~Myr later ($z\simeq6$), when star formation has been quenched. The black horizontal lines show the sensitivities of JWST/NIRCam filters for point sources with S/N=5 and exposure time of $\sim3$~hrs. The dashed curves show the  intrinsic (i.e. unattenuated) stellar emission for three satellites.
\label{fig:sed}
}
\end{figure*}

The intrinsic spectrum can be attenuated by dust, for which we adopt the SMC synthetic extinction curve by \citet{Weingartner01}. 
 This is the natural choice for dwarf galaxies and a good approximation for \highz sources at $\lambda > 0.3\, \micron$ \citep{Gallerani2010}.
We compute the optical depth as: $\tau = C_{\rm ext} \times N_{\rm H} [\cm^{-2}] \times Z [\zsun]$, where $C_{\rm ext}$ is the tabulated cross section per H nucleon, and $N_{\rm H}$ and $Z$ are the gas hydrogen column density and metallicity averaged on different line-of-sights. 
The latter two are retrieved directly from the simulation, considering the gas composition in the selected satellite galaxy at the snapshots available (i.e. every 18 \myr), and interpolating at intermediate times.
The final SEDs (Fig.~\ref{fig:sed}) are obtained by reprocessing the intrinsic stellar spectra with the derived dust transmission curves.

\vspace{0.05in}

\begin{figure*}
\centering
\includegraphics[width=0.95\textwidth]{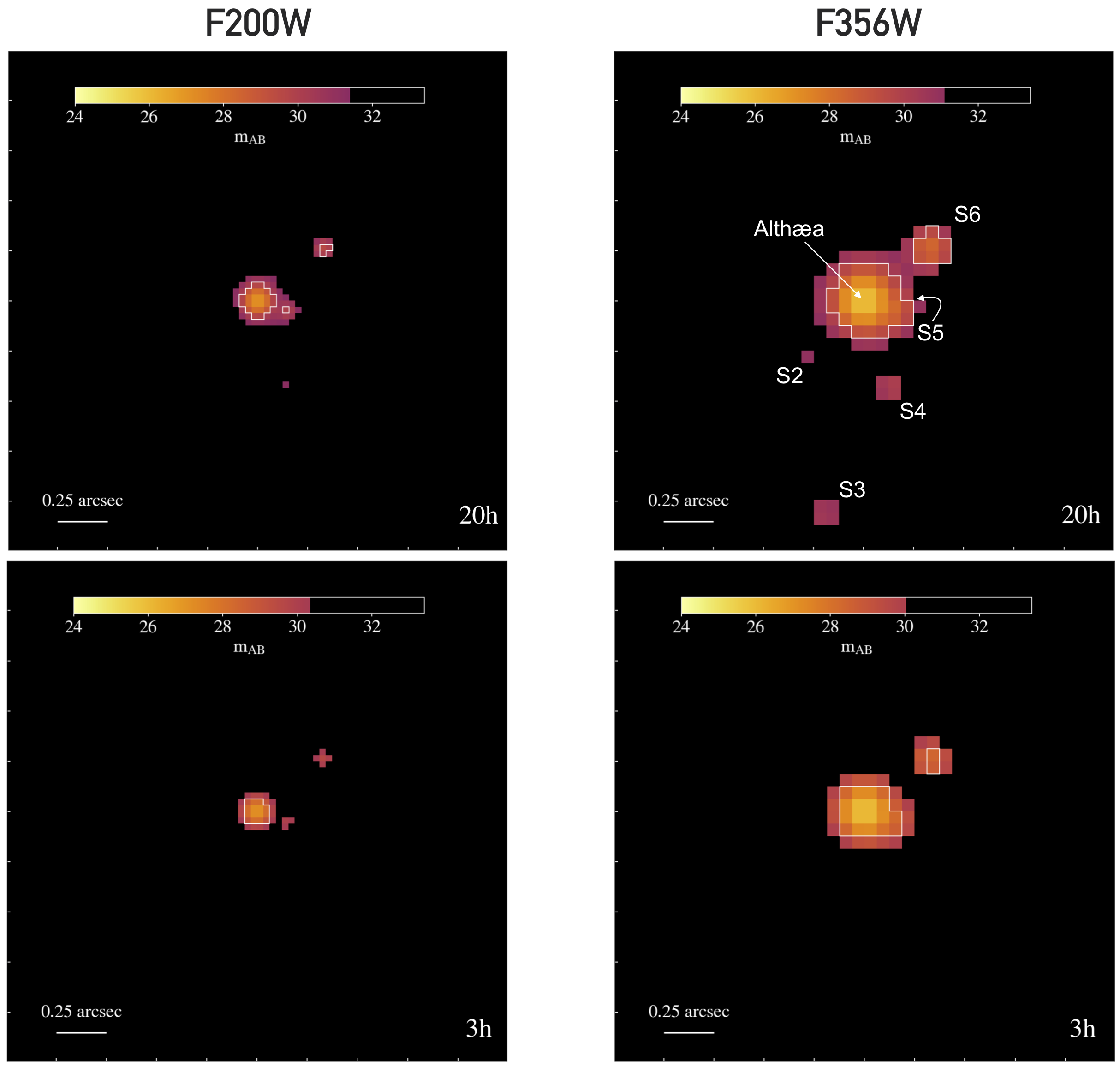}
\caption{Synthetic images centred on the LBG at $z=6$ in a field of view of size $2.5 \times 2.5$~arcsec, corresponding to $15 \times 15$~kpc. The maps are displayed in two NIRCam filters: F200W (left, short-wavelength channel with pixel scale of $0.032\arcsec$) and F356W (right, long-wavelength channel with pixel scale of $0.065\arcsec$), and for exposure times of 20 hours (top) and 3 hours (bottom). The images are shown for S/N=3 and the white contours identify the pixels with S/N=10. The location of the satellites are pinpointed by the labels S2,...,S6 from the least to the most massive system.
\label{fig:image}
}
\end{figure*}

To enable a detailed spatial/morphological analysis, we have produced synthetic images of the central LBG and its dwarf satellites at $z\sim6$ (Fig.~\ref{fig:image}). The images are built from stellar density maps of the simulated, face-on LBG disk \citep[e.g. see Fig.~2 in][]{Gelli20}, and using the proportionality between stellar mass and observed magnitude in the selected bands. We simulate the instrument effects by convolving the high-resolution images with a gaussian with the FWHM of the selected NIRCam filter, and then rebinning to match the instrument pixel size. No sky background has been simulated when creating the images.

\section{Results}%%%%%%%%%%%%%%%%%%%%%%%%%%%%%%%%%%%%%%

The resulting SEDs of the satellites and the central LBG \althaea~are shown in Fig.~\ref{fig:sed}, in the NIRCam wavelength range, $0.6-5\, \micron$, and for two different redshifts: $z\simeq6.5$ (left panel) and $z\simeq6$ (right panel). We adopt the same notation as in \citet{Gelli20} to indicate the satellites, i.e. they are named from S1 to S6 with increasing stellar mass. S1 is the proto-globular cluster: its flux is more than 2 order of magnitudes below that of the other satellites, hence invisible even for JWST deep surveys.
We notice that the flux of the central LBG (gray dashed) always exceeds that of the satellites by more than an order of magnitude at $\lambda > 2\micron$, because of the larger stellar mass. The two displayed redshifts pin-point two relevant evolutionary stages of the 3 smallest satellites (S2, S3, S4, $M_\star \lesssim 5\times 10^8 \msun$):
\begin{itemize}
\item[$-$] \textit{Bursty phase} ($z\simeq 6.5$, left panel): in this epoch the least massive satellites are all experiencing the brief ($\lesssim~50~\myr$) burst of star formation in which they form all of their stars. At this stage they are characterised by relatively high star formation rates, up to ${\rm {\rm {\rm SFR}}}\sim14~\msunyr$. Their stellar populations are hence dominated by young newly formed stars, resulting in spectra with an intense rest-frame UV flux at $\lambda\lesssim2.5\micron$.

\item[$-$] \textit{Passive phase} ($z\simeq 6$, right panel): this stage takes places $\sim100~\myr$ later, when the star formation in the least massive satellites has been completely suppressed by SN feedback. The older stars and lack of ongoing star formation result in redder spectra. Note that at $\lambda\lesssim2.5\micron$ the flux has decreased by a factor $\sim 10$.
\end{itemize}

SEDs of the two most massive satellites (S5 and S6, $M_\star \gtrsim 5\times 10^8 \msun$) are dominated by old stellar populations in both epochs, since they are older than 300~Myr. The SFH of these systems is characterised by a low-level, continuous activity fuelled by merger events which produce peaks with ${\rm {\rm {\rm SFR}}}<5 \msunyr$ \citep[e.g. see Fig.~7 from][]{Gelli20}.
By $z\sim6.5$ these high-mass systems can be therefore considered as ``stabilised'' since the bulk of their stellar population is already in place. As a result, their SED is almost unchanged between the two stages.
Also, in the first stage they are fainter than S2, S3, S4 due to the lower {{\rm SFR}}, and to their more dusty nature; as lower systems fade, massive systems are the brightest ones in the passive phase\footnote{The ${\rm {\rm SFR}}$ of the most massive satellite (red curve) is temporarily zero during the second stage at $z\simeq6$ because of a recent ($\sim 10~\myr$ before) burst of star formation and the subsequent SN feedback effect.}.

To highlight dust effects on the final spectra, we compare the intrinsic stellar emission for S3, S4 and S6. As the dust content broadly scales with galaxy stellar mass, massive systems suffer a larger attenuation. 
For instance, even though S3 and S4 have similar intrinsic fluxes at wavelengths $\lambda\sim1\micron$ during the bursty phase, the more massive S4 is more attenuated. Low-mass systems are, however, particularly affected by dust extinction during their starburst phase, which is characterized by higher gas column density in the star forming regions.
Interestingly, the predicted fluxes are in most cases well above the sensitivity of NIRCam in different filters, shown for a $\rm S/N=5$ in $\sim3$~hrs by horizontal black lines. However, these estimates are valid for point sources. To draw solid conclusions on their observability, we need to assess whether the satellites can be spatially resolved by JWST.

\vspace{0.05in}

Fig.~\ref{fig:image} shows the synthetic images of the LBG system at $z\sim6$ over an area of $2.5\arcsec\times2.5\arcsec$ in two wide filters (F200W and F356W). The coloured pixels within the images show the flux at S/N$\sim3$, with exposure times of $\sim20$~hrs (top) and $\sim3$~hrs (bottom). The white contours indicate\footnote{Flux limits for $\sim3$~hrs are retrieved from Table1 of \textit{https://jwst-docs.stsci.edu/near-infrared-camera/nircam-predicted-performance/nircam-sensitivity}, which assumes circular photometric apertures 2.5~pixels in radius and 1.2~$\times$~minimum zodiacal light background. We rescale these values for 20 hrs, consistently with the values reported in \citet{Rieke19}.}
 the areas with S/N$\sim10$.
The image shows the system at $z\sim6$, i.e. the last snapshot available during the passive phase, when the three low-mass satellites are at their faintest level. This conservative choice reveals how, even in this worst case of detectability conditions during their evolution, many satellites do exceed the sensitivity threshold. Namely, the fluxes of S5 and S6 reach S/N=10 in both filters and they are detectable in just 3~hrs of observations.
The emission of all 5 dwarf galaxy satellites exceeds the S/N=3 sensitivity threshold in the wide filter F356W in 20 hrs. In F200W, S2 and S3 remain undetected at $z\sim6$, but they are detectable in $\sim 17$~hrs at previous stages when their flux is higher ($6.5\lesssim z\lesssim 6.2$).
Fig.~\ref{fig:image} shows that to distinguish the dwarf galaxy satellites from the central LBG, a separation of $\gtrsim 0.25\arcsec$ is required. Noticeably, this requirement is achieved for all dwarf satellites with the exception of S5, which is located in the inner regions ($1.2~\kpc$ from the centre of \althaea) and whose emission is likely confused with that of the central LBG. 

\vspace{0.05in}
\begin{figure*}
\centering
\includegraphics[width=0.9\textwidth]{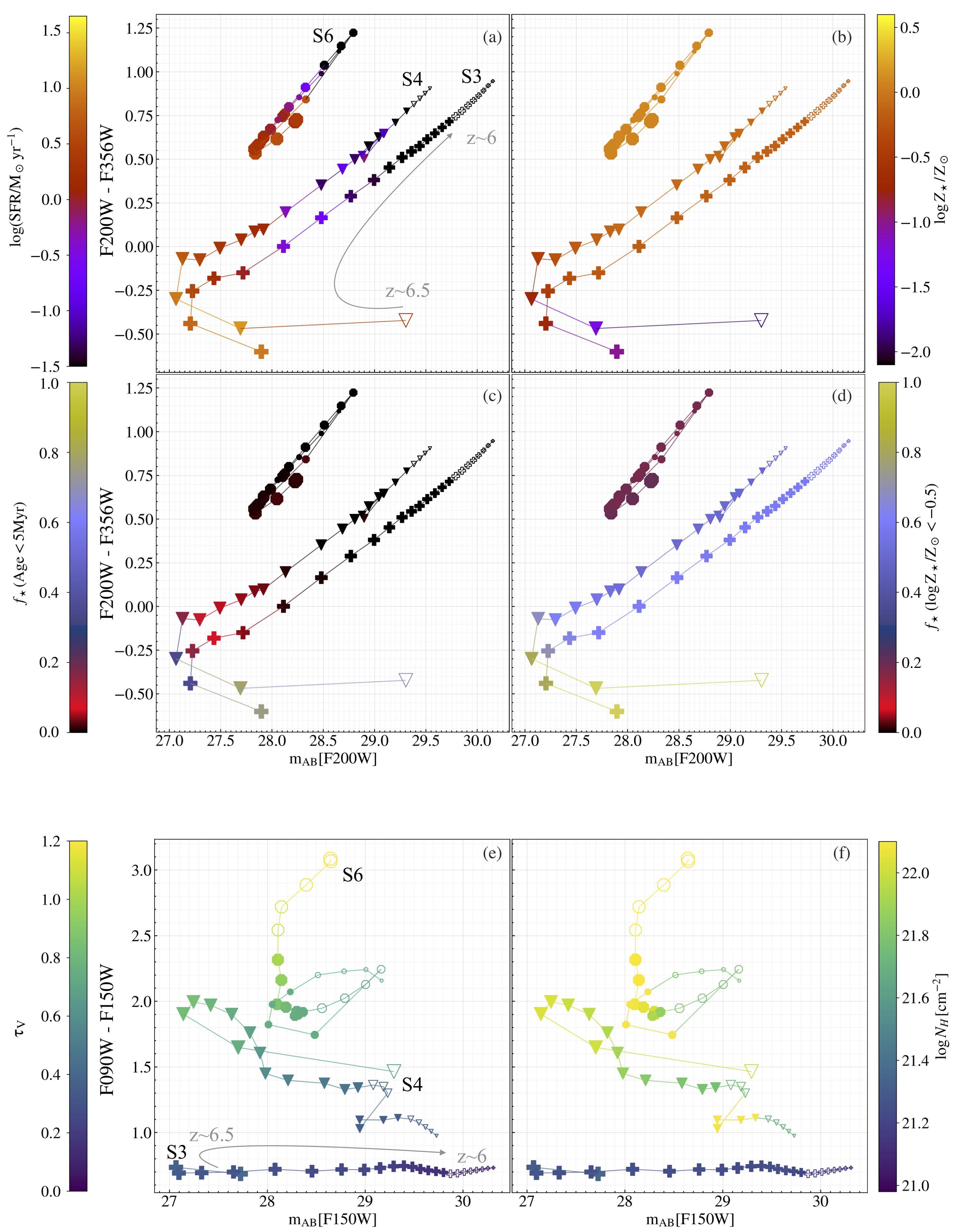}
\caption{Color-magnitude diagrams of three satellites using different NIRCam filters. A line traces the evolution of each satellite in the diagram, the size of the markers decreasing in cosmic time from $t\sim850~\myr$ (in correspondence of the bursty phase of low-mass satellites) to $\sim950~\myr$ (passive phase), in steps of $5~\myr$. Empty markers pinpoint the stages at which the satellite do not exceed the S/N=3 sensitivity threshold in 20 hrs in the selected filters.
In each panel the markers are colour-coded according to different quantities. The first four characterise stellar populations: ${\rm {\rm SFR}}$, stellar metallicity, fraction of young stars ($\rm Age<5~\myr$), and fraction of metal poor stars [$\log Z_\star/\zsun < -0.5$]. The last two refer to the ISM, i.e the dust optical depth $\tau_{\rm V}$, and hydrogen column density $N_{\rm H}$.
\label{fig:colors} 
}
\end{figure*}

In conclusion, dwarf satellites of high-z LBGs can be actually observed with JWST within the planned deep surveys exceeding $\sim20$~hrs of exposure time.
What information can be inferred from their photometry, in particular through colour-magnitude diagrams (CMDs)? While retrieving colours and magnitudes, we checked the possible contamination by nebular emission lines (e.g. $\rm H_\alpha$, $\rm H_\beta$, [OIII]$\lambda$5007 and [OII]$\lambda$3727), post-processing the simulation outputs with \code{cloudy} \citep{Ferland17}, with the caveat of not having a consistent modeling of the interstellar radiation field with radiative transfer \citep[see][]{Pallottini19}. Our estimates show that emission lines are expected to produce negligible variation in both colour and magnitude ($<0.1$dex) for the three satellites farther from \althaea~ (S3, S4, S6), while the two located in its proximity ($\lesssim0.4\arcsec$) suffer heavier line contamination due to the nearby LBG disk.
Fig.~\ref{fig:colors} shows the CMDs from $z\sim6.5$ ($t_{Uni}\sim850~\myr$) to $z\sim6$ ($t_{Uni}\sim950~\myr$) of the 3 best-target satellites which (i) are located farther from the central LBG, and (ii) can be identified in only 7 hrs observations in F356W at $z\sim6$ (the faintest evolutionary stage).

In the upper four panels (a-b-c-d) we plot the color F200W-F356W as a function of the AB magnitude in the filter F200W, and the symbols are color-coded according to ${\rm {\rm SFR}}$, stellar metallicity, $Z_\star$, fraction of young, $f_\star(\rm Age<5~\myr$), and metal-poor $f_\star(\rm Z_\star < 10^{-0.5}\zsun)$ stars. 
The path of each satellite in the CMD helps clarify how its SED evolves with time. 
We notice two trends: (i) a complex one characterising  ``active'' low-mass satellites (S3 and S4), which first ({\it bursty phase}) become brighter (from $\rm m_{AB}\approx29$ to $\rm m_{AB}\approx27$) with roughly constant colour (F200W-F356W~$\approx-0.5$), and then ({\it passive phase}) become progressively dimmer and redder as they age; (ii) the simpler one of the high-mass S6, which by $z\approx 6.5$ is already ``stabilized'', i.e. old/red (F200W-F356W$\gtrsim0.5$), and therefore only shows a modest variation ($\lesssim 1$ dex) in both magnitude and colour.

Since the high-mass satellite forms only $\sim6\%$ of its stellar mass during the time considered, no significant variation takes place in terms of both $Z_\star$ (always super-solar, panel b) and $ f_\star(\rm Z_\star < 10^{-0.5}\zsun)$ ($\lsim15\%$, panel d). Conversely, its ${\rm {\rm SFR}}$ varies due to a short bursts of star formation (${\rm {\rm SFR}}\sim5\msunyr$, panel a) 
which is however too weak to significantly affect the fraction of young stars (panel c). 
On the other hand, low-mass satellites, during the first $\sim10~\myr$ displayed, experience the peak of their bursty star formation period in which they form the bulk of their stellar mass. The ${{\rm SFR}}$ reaches its peak ($\gtrsim10~\msunyr$, panel a), and the SED is mostly determined by newly formed young ($\gtrsim50\%$, panel c), and metal-poor stars ($ \gtrsim80\%$, panel d). The stellar metallicity shows a rapid increase during the bursty phase ($Z_\star\gtrsim10^{-1}\zsun$, panel b).
Over the $\sim100~\myr$ shown, these low-mass systems spend only $\sim20~\myr$ in the initial bursty phase, implying that in principle, when observing a \highz low-mass satellite, it will be more likely to detect it during the longer passive phase when their ${\rm {\rm SFR}}$ has decreased and $Z_\star$, $f_\star(\rm Age<5~\myr$), and $ f_\star(\rm Z_\star < 10^{-0.5}\zsun)$ settle to their final, constant values.

The two bottom panels (e-f) show the color F090W-F150W as a function of the F150W magnitude, and are colour-coded according to the V-band dust optical depth, $\tau_{\rm V}$, and $N_{\rm H}$. 
The SED shape in this spectral range is determined by two competing factors: young stellar populations, enhancing the flux at short wavelengths, and dust drastically reducing it.
The trends in the CMD show that in general more massive satellites, which are dustier and denser, have higher F090W-F150W.
The smallest object displayed (S3) shows a constant trend of the F090W-F150W, reflecting its simple SFH in which it settles to constant values of metallicity and column density. More massive satellites show a more complex behaviour, in which the color becomes bluer at each small burst episode. The trends of the optical depth (e) and column density (f) are associated with the color: for instance, objects with F090W-F150W~$\gtrsim2.5$ have $\tau_{\rm V}\gtrsim1$ and $N_{\rm H}\gtrsim10^{22}\rm cm^{-2}$.

\section{Discussion}%%%%%%%%%%%%%%%%%%%%%%%%%%%%%%%%%%%%%%

Based on our high-resolution cosmological simulations, we show that JWST will be able to detect for the first time faint dwarfs satellite galaxies of typical LBGs at the end of the Epoch of Reionization. Remarkably, this should happen already during the first planned observations. The JADES survey, for instance, will reach exposure times of $\sim20$~hrs, sufficient for observing most satellites even during the faintest stages of their evolution.
The survey will detect $\sim 50$ LBGs with stellar mass $M_\star\sim10^{10}\msun$ \citep{Williams18}, i.e similar to \althaea. According to our simulations, we then expect to detect a total of $\sim100-250$ companion dwarf satellites, $\sim40$ of which will be most likely star bursting and dominated by young and metal-poor stars. In a future work (Gelli et al. in prep) we will investigate the expected statistics of \highz dwarf satellite galaxies and probability of observing them in different phases, using a larger sample of LBG targets from the \code{SERRA} suite (Pallottini et al. in prep.).
The simulation used in this work \citep{Pallottini17} is ideal to study this kind of sourced as it reproduces key observed properties of known LBGs, from $L_{\rm [CII]}$ luminosities \citep{Carniani18} to the Schmidt-Kennicutt relation \citep{Krumholz09}. 

Once candidate dwarf satellites are individuated based on the proximity to a massive LBGs, a measurement of their photometric redshift is needed to infer if they do belong to the LBG system, or if we are rather dealing with an interloper. NIRCam will be able to retreive $z\sim6$ galaxies redshifts with a percentage of misidentifications as lower $z$ galaxies of $9\%$ \citep{Bisigello16}, hence allowing a successful identification of the target LBG satellites.

An additional interesting finding of the simulation analysis \citep[see][]{Gelli20} is the presence of a proto-globular cluster among the satellite population of the LBG at $z\sim6$. This is characterised by a stellar mass $M_\star\sim10^6\msun$, formed in an single burst of star-formation ($<15~\myr$), and completely lacking both gas and dark matter. In this study, we find that such objects, despite their predicted existence in the vicinity of LBGs, are not bright enough to be observed with JWST/NIRCam: more than $\sim1000$~hrs would be required to detect their light at S/N$\sim3$. Since within the JADES survey we do not expect the presence of gravitationally lensed sources, the hundreds of expected observed satellites around LBGs will be indeed dwarf galaxies, and not globular clusters \citep[e.g.][]{Vanzella17}. Among them, some will be most likely detected while experiencing their burst of star formation. 
We showed how photometric observations will be most relevant to shed the first light on \highz satellite galaxies and, provided that many of these sources will be discovered in the surroundings of LBGs, we
suggested that the F200W-F356W color is a powerful diagnostic tool to understand their properties. 
For example, F200W-F356W~$\lesssim-0.25$ characterises star-bursting (${\rm SFR}\sim~5~\msunyr$), low-mass ($M_\star\lesssim5\times 10^8\rm \msun$), low stellar metallicity systems, with $\sim80\%$ of their stellar population composed by young and metal-poor [$\log(Z_\star/Z_\odot) < -0.5$] stars.

%\begin{acknowledgments}
\vspace{0.25in}
We thank the referee for the careful reading of the paper and the insightful comments.
We acknowledge support from PRIN-MIUR2017, prot. n.2017T4ARJ5, ERC Starting Grant NEFERTITI H2020/808240 and ERC Advanced Grant INTERSTELLAR H2020/740120.
%\end{acknowledgments}

\bibliography{LBG_satellites_with_JWST}
\bibliographystyle{aasjournal}

\end{document}